\documentclass[nat,twocolumn,superscriptaddress,showpacs]{revtex4-1}

\usepackage{natbib}
\usepackage{setspace}
\usepackage{textgreek}
\usepackage[dvipsnames]{xcolor}
\usepackage{amsmath}
\usepackage{enumitem}
\usepackage{ragged2e}
\usepackage{float}
\usepackage{graphicx}
\usepackage{xcolor}
\usepackage{dcolumn}
\usepackage{bm}
\usepackage[utf8]{inputenc}
\usepackage{bbm}

\usepackage{hyperref}
\usepackage{nicefrac}
\usepackage{color}
\usepackage{braket}
\usepackage{multirow}
\usepackage{verbatim}
\usepackage{bbold}

\usepackage{braket}

\begin{document}

\title{Edge states in proximitized graphene ribbons and flakes in a perpendicular magnetic field: emergence of lone pseudohelical pairs and pure spin-current states}

\author{\surname{Yaroslav} Zhumagulov}
\email[Emails to: ]{iaroslav.zhumagulov@ur.de}
\author{\surname{Tobias} Frank}
\author{\surname{Jaroslav} Fabian}
\email[Emails to: ]{jaroslav.fabian@ur.de}

\affiliation{%
 Institute for Theoretical Physics, University of Regensburg,\\
 93040 Regensburg, Germany
 }%

\date{\today}

\begin{abstract}
We investigate the formation of edge states in graphene ribbons and flakes with proximity induced valley-Zeeman and Rashba spin-orbit couplings in the presence of a perpendicular magnetic field $B$. Two types of edges states appear in the spin-orbit gap at the Fermi level at zero field: strongly localized pseudohelical (intervalley) states and weakly localized intravalley states. We show that if the magnetic field is stronger than a crossover field $B_c$, which is a few mT for realistic systems such as graphene/WSe$_2$, only the pseudohelical edge states remain in zigzag graphene ribbons; the intravalley states disappear. 
The crossover is directly related to the closing and reopening of the bulk gap formed between nonzero Landau levels.
Remarkably, in finite flakes the pseudohelical states undergo perfect reflection at the armchair edges if $B > B_c$, forming standing waves at the zigzag edges. These standing waves comprise two counterpropagating pseudohelical states, so while they carry no charge current, they do carry (pure) spin current.

\end{abstract}

\maketitle

\section{Introduction}
Modification and external control of the electronic structure of two-dimensional materials via the proximity effect is of great experimental and technological interest for designing systems with novel magnetic and spin properties~\cite{Han2014,Sierra2021,RevModPhys.92.021003}. Of particular importance is the introduction of 
spin interactions in graphene, whose Dirac electrons exhibit only weak spin-orbit coupling. Placing graphene in proximity to transition metal dichalcogenides (TMDCs) is a viable route, resulting in spin-orbit couplings on the meV scale~\cite{Gmitra2015,Wang2015, Gmitra2016, Gani2020},
also tunable by twisting \cite{Li2019,David2019,naimer2021twistangle,pezo2021manipulation}. While the proximity spin-orbit coupling is in general sublattice dependent \cite{Frank2016, Kochan2017}, 
TMDC substrates induce valley-Zeeman and Rashba couplings, providing uniform pseudospin spin-orbit fields, opposite at $K$ and $K'$~\cite{Gmitra2015, Wang2015,PhysRevB.94.241106}. 
Similar proximity spin-orbit physics and the appearance of the valley-Zeeman coupling has been predicted for graphene on Bi$_2$Se$_3$-family of topological insulators \cite{Song2018,PhysRevB.100.165141}, but also for 
bilayer Jacutingaite\cite{Rademaker2021}. There already exists a significant body of experimental evidence demonstrating the presence of valley-Zeeman coupling in proximitized graphene~\cite{Garcia2018,Island2019,Karpiak2019,Tang2020,Hoque2021,Ghiasi2017,Benitez2018,Ghiasi2019,Safeer2019,Herling2020,Bentez2020,Avsar2014,Banszerus2020,Khokhriakov2020, Wakamura2019, Wakamura2020,amann2021gatetunable,2021boosting,InglaAyns2021}.

In proximitized graphene with valley-Zeeman coupling, two groups of edge states---pseudohelical (intervelley) and intravalley states---form within the spin-orbit gap. At each edge there are those two pairs of states, conforming to the trivial topology of the system \cite{Frank2018}. However, for ribbons on the nanoscale, with the widths less than a micron, the intravalley states are gapped out by the confinement-induced hybridization and only the lone pseudohelical pair remains at each edge. This pair is fully protected against backscattering by time-reversal defects, similarly to helical states of the spin quantum Hall effect \cite{Kane2005}. But, unlike helical states, the pseudohelical states in a flake change the helicity from one sigzag edge to the other, flipping the spin along the armchair edge at which perfect tunneling of the states occurs. In large flakes, where also intravalley states propagate, the backscattering protection by time-reversal symmetry is lifted, as the pseudohelical states reflect at armchair edges to intravalley states at the same edge. 

Is there a way to reinstate the lone pseudopotential pair also in larger ribbons and flakes, where there are nominally intravelley states as well. We show in this paper that placing proximitized graphene in a perpendicular magnetic field achieves exactly that. Magnetic effects are typicall twofold: Zeeman-like, providing spin imbalance, and orbital effects, leading to Landau quantization. Zeeman effects in proximitized graphene were investigated earlier, demonstrating the appearance of the quantum anomalous Hall effect \cite{Hogl2020} and chiral Majorana modes \cite{Hogl2020a}. Magnetic orbital response of helical edge states to magnetic fields was studied in previous works \cite{Luo2020,Frank2020,Yang2016,Li2018,DeMartino2011,Lado2015,Delplace2010,Gani2020}. Rather surprisingly, the quantum spin Hall edge states, which are generated by uniform intrinsic (Kane-Mele) spin-orbit coupling, are not necessarily destroyed by the cyclotron effect~\cite{Shevtsov2012, Goldman2012}, which can theoretically be used to switch between the quantum spin Hall and quantum Hall regimes by gating. However, there can be a crossover between topological and trivial regimes in the presence of a perpendicular magnetic field  ~\cite{Scharf2015,Bottcher2019}. 

\begin{figure}
\begin{center}
\includegraphics[width=.5\textwidth]{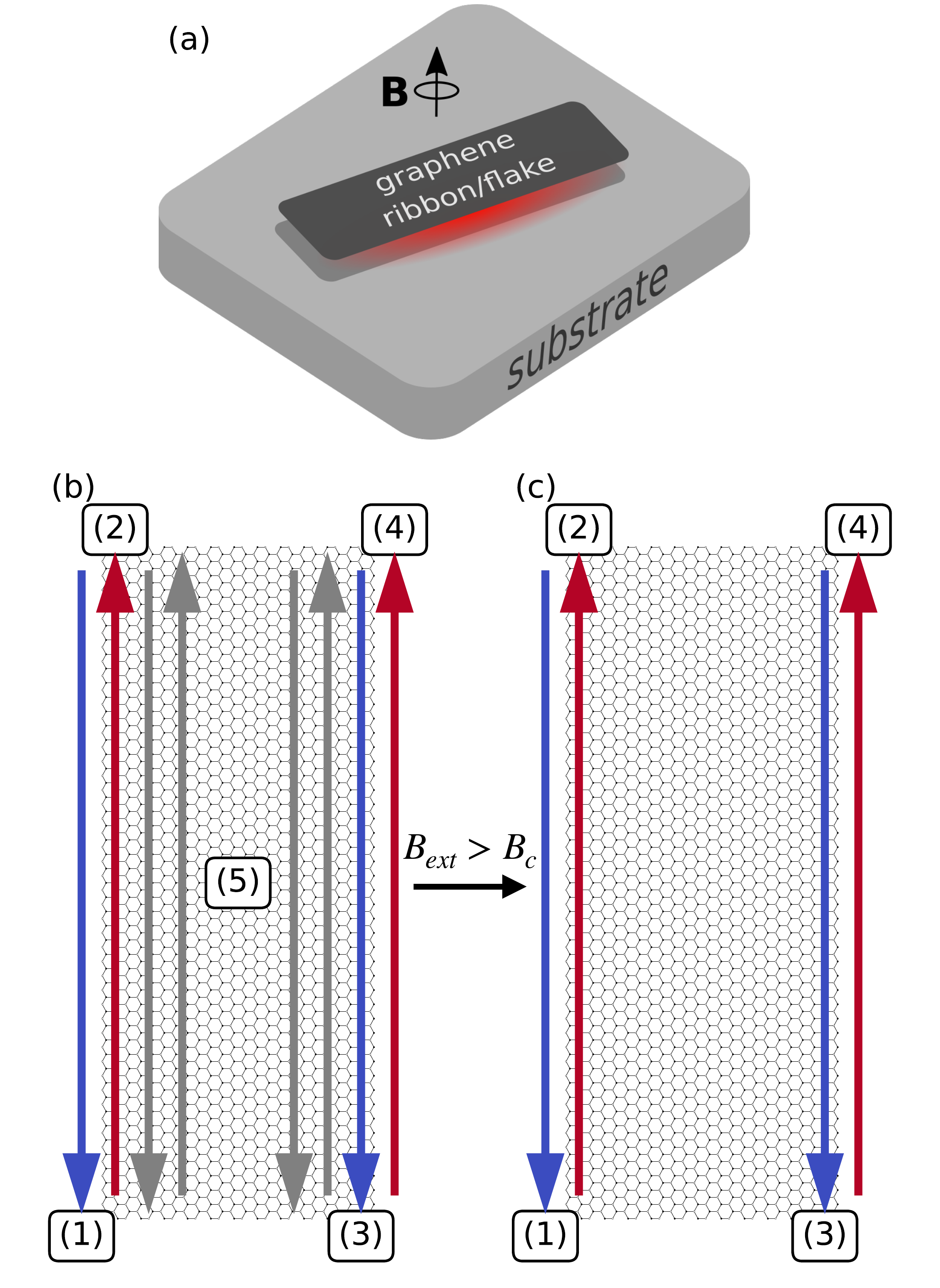}
\end{center}
	\caption{(a) Sketch of a graphene ribbon or flake on a substrate such as a TMDC which yields valley-Zeeman spin-orbit and Rashba couplings. A perpendicular magnetic field is applied to affect the orbital states of the Dirac electrons. (b) Two types of edge states near the Fermi level in graphene with valley-Zeeman spin-orbit interaction: (1-4) spin-polarized pseudohelical (intervalley) states, and (5) intravalley states. (c) Lone pairs of pseudohelical states above the crossover magnetic field.}
\label{fig:schema}
\end{figure}

In this paper we study theoretically the response of the pseudohelical and intravalley edge states in proximitized graphene (using realistic paramaters for a graphene/WSe2 heterostucture) to an external perpendicular magnetic field, as depicted in Fig. 1.(a). We employ a tight-binding model supplemented with Peierl's substitution to study the electronic structure of graphene zigzag nanoribbons and finite flakes. The Landau levels calculated by the tight-binding approach are in excellent agreement with the bulk Landau level predictions \cite{Frank2020}. The pairs of pseudohelical  edge states, see Fig.\ref{fig:schema}(b), are preserved even if time-reversal symmetry is broken by the magnetic field. 
This is 
similar to what happens in the quantum spin Hall effect, where helical edge states are preserved in an applied field~\cite{Shevtsov2012}.

However, the  intravalley states (originating from Rashba SOC~\cite{Frank2018}) disappear once the magnetic field increases beyond some critical value $B_c$. At $B > B_c$, intravalley edge states merge with the conduction and valence bands, opening an intravalley gap. 
Inside the gap one finds a lone pair of pseudohelical states, at each zigzag edge. Effectively, the magnetic field gaps out the weakly localized intravalley states, mimicking finite-size confinement. \cite{Frank2018}. 

The paper is organized as follows. In Sec. II. we introduce a tight-binding model Hamiltonian for graphene with proximity induced spin-orbit coupling, and present a scaling method to obtain the energy spectrum for microscopic structures. In Sec. III. we analyze the electronic structures of zigzag ribbons and finite flakes states in the presence of an external perpendicular magnetic field. We also discuss the crossover at which the intravalley edges states are gapped, referring to bulk Landau level results. Finally, in Sec. IV., we briefly summarize the main results.

\section{Model and Methods}

We consider Dirac electrons in proximitized graphene, with sizeable (on the meV scale) spin-orbit interactions of the valley-Zeeman and Rashba types. Such interactions are induced by the proximity effects with TMDCs or TIs, as discussed above. To investigate finite systems, zigzag ribbons and flakes, we implement the following tight-binding  Hamiltonian, \cite{Gmitra2016,Kochan2017,Gmitra2013}
\begin{align}
    \hat{\mathcal{H}}&=\sum_{\langle i,j \rangle}t c_{is}^\dagger  c_{js}+\sum_{i}\Delta\xi_i c_{is}^\dagger  c_{is} \nonumber \\
    &+\frac{2i}{3}\sum_{\langle i,j \rangle}\lambda_R c_{is}^\dagger  c_{js^\prime}\left[\left(\hat{\mathbf{s}}\times \mathbf{d}_{ij}\right)_{z}\right]_{ss^\prime} \nonumber \\
    &+\frac{i}{3}\sum_{\langle\langle i,j \rangle\rangle}\frac{\lambda_{I}^{i}}{\sqrt{3}}c_{is}^\dagger c_{js} \left[\nu_{ij}\hat{\mathbf{s}}_z\right]~,  \label{eq:tb-ham}
\end{align}
where $c_{is}^\dagger$ and $c_{is}$ are the creation and annihilation operators for site $i$ and spin $s$, $\langle i,j \rangle$ denotes the nearest, $\langle\langle i,j \rangle\rangle$ the next nearest neighbors.
The Hamiltonian of Eq.~\eqref{eq:tb-ham} has four terms: First, the nearest-neighbour hopping with amplitude $t$ occurs between sites $i$ and $j$ with spin preservation. Second, the proximity effects induces the staggered potential $\Delta$ with signs $\xi_i$ equal to $+1$ or $-1$ for A and B sublattices, respectively. The third term describes
Rashba spin-orbit coupling \cite{Rashba2009,Tsaran2014} with amplitude $\lambda_{R}$, which breaks horizontal reflection symmetry and mixes states of opposite spins and sublattices. Symbols $\mathbf{d}_{ij}$ and $\hat{\mathbf{s}}$ denote the unit vector from site $j$ to site $i$ and the vector of spin Pauli matrices, respectively. Finally, the fourth term models the  valley-Zeeman spin-orbit coupling \cite{Gmitra2015, Wang2015, Yang2016}. This term preserves the spin, but the intra-sublattice hopping is different for  clockwise ($\nu_{ij} = -1$) and counterclockwise ($\nu_{ij} = +1$) paths along a hexagonal ring from site $j$ to $i$. The intrinsic spin-orbit coupling $\lambda_{I}^{i}$ is written here in a more general way, allowing for different strengths at $A$ and $B$ sublattices. The orbital effects of a perpendicular magnetic field are modelled by Peierl's phase \cite{Peierls1933,PhysRev.115.1460}. We do not consider
the Zeeman effects of the field, as they are negligible for the fields we consider, on the millitesla scale. 

In the following, when we present numerical results, we use parameters from first-principles
calculations for graphene/WSe$_2$, see Ref. \cite{Gmitra2016}: the nearest-neighbor hopping $t=-2.507$~eV, staggered potential $\Delta=0.56$~meV, Rashba SOC parameter $\lambda_R=0.54$~meV, and intrinsic SOC parameters $\lambda_I^A=1.22$~meV and $\lambda_I^B=-1.16$~meV. 

We wish to investigate micron-size systems, in which magnetic orbital effects are not quenched, and in which the intravalley edge states are not gapped out. Since the spin-orbit parameters are on the meV scales, the considered systems need to be large enough to resolve such energies in the subband structure. To reduce computational efforts and avoid dealing with intractably large structures, we employ a scaling trick ~\cite{Liu2015}, which allows us to consider smaller structures but with rescaled parameters. 
Finite-size level spacings in graphene ribbons are on the order of $\Delta E\approx\pi\hbar v_F / w$~\cite{Lin2008}, which can be reduced by changing the Fermi velocity $v_F\rightarrow v_F/r= \sqrt{3}a t /2\hbar r$, by rescaling the nearest-neighbor hopping $t$. Such a rescaling affects the interpretation of the lattice constant $a$, which is mapped to $ra$, in order to keep the energy spectrum invariant. The cyclotron energy in graphene is given by $\hbar\omega_c = \sqrt{2e\hbar B} v_F$. To preserve this energy scale, we rescale the external perpendicular magnetic field by $B\rightarrow Br^2$. The rescaling does not influence the underlying physics as long as we consider low-energy states for which the linear dispersion holds\cite{Liu2015}. 

The scaling trick permits us to keep the cyclotron energy scale at the level of SOC parameters, while decreasing the computational burden to simulate large-scale systems. The main criterion for choosing the rescaling parameter is $r\ll3t\pi/E_{\text{max}}$, where $E_{\text{max}}$ is the maximal energy of interest~\cite{Liu2015}. We wish to resolve energies up to $E_{\text{max}}=2$ meV, which safely covers the spin-orbit gap region in which the edge states form, the scaling parameter  $r\ll3t\pi/E_{\text{max}}$=11809. In our work we we choose $r=400$. With that, the finite-size effects are decreased to $0.05$~meV for a width of 400 unit cells. The magnetic field strength equivalent to this energy scale is about $10^{-5}$~T, well below the value for the crossover magnetic field (see the next chapter) for gap closing.

To calculate the band structure of zigzag nanoribbons and the electronic states of a graphene flake with proximity induced spin-orbit interaction, we implement the above model using the Python-based numerical package KWANT~\cite{Groth2014}.

\section{Results}

\subsection{Bulk results}

The starting point for analyzing the behavior of edge states of proximitized graphene in the presence of an external perpendicular magnetic field is the study of Landau levels in the bulk system. A detailed derivation of the Hamiltonian of the Landau levels for bulk proximitized graphene is presented in Ref.~\cite{Frank2020}. Graphene with spin-orbit coupling in magnetic fields was studied from different perspectives by other groups as well~\cite{Cysne2018, DeMartino2011}.

\begin{figure}
\begin{center}
\includegraphics[width=.5\textwidth]{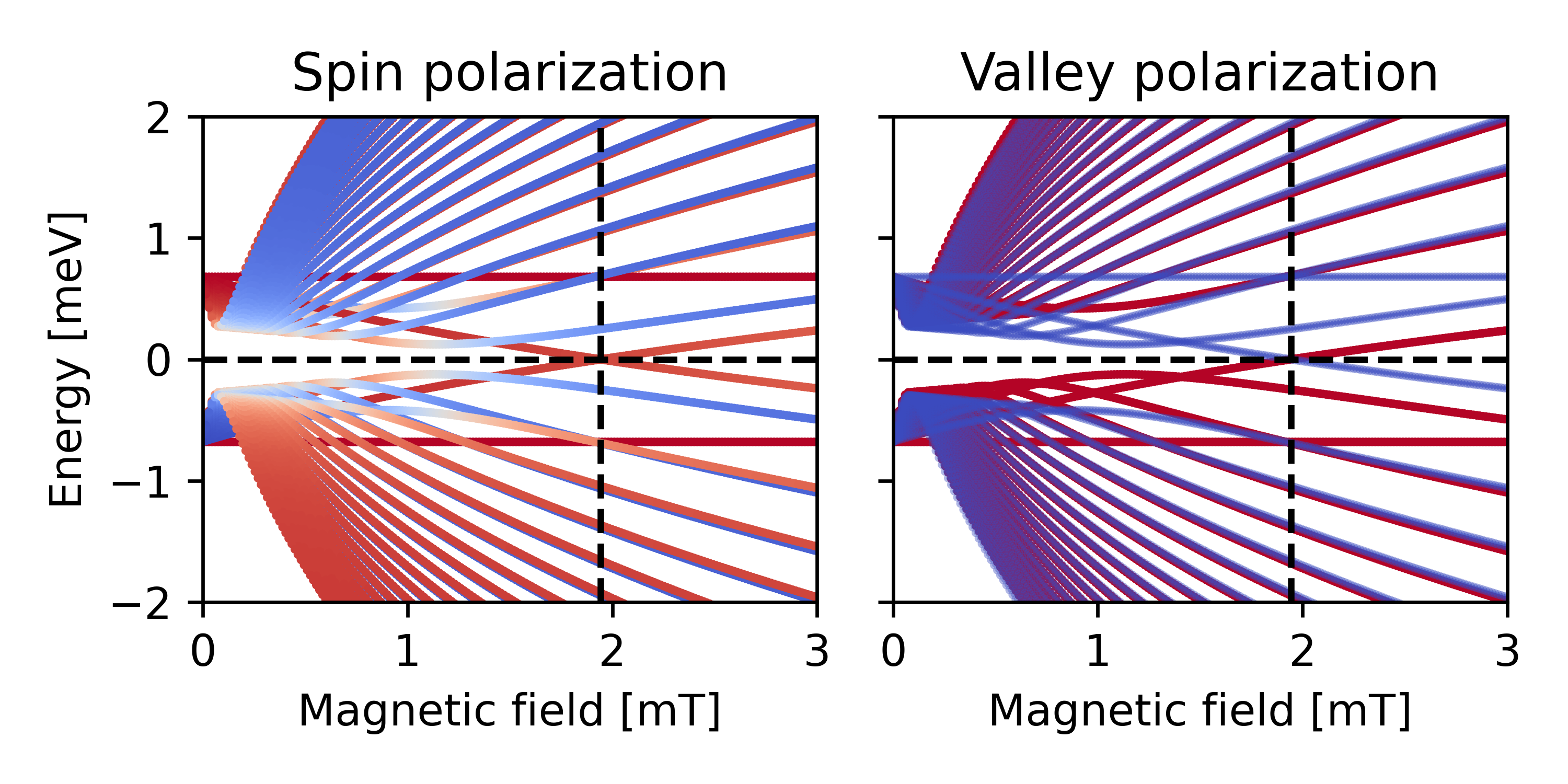}
\end{center}
\label{fig:bulk_landau}
\caption{Calculated evolution of the bulk Landau levels in graphene/WSe$_2$ with increasing of the external magnetic field. The crossover magnetic field is indicated as a dashed line. The color code corresponds to the $s_z$ expectation value in the left column and $\kappa_z$ (valley) expectation value in the right column, see Ref.~\onlinecite{Frank2020}.}
\end{figure}

We construct the Landau fan diagram, shown in Fig.~\ref{fig:bulk_landau}. Comparing to the Kane-Mele model \cite{Kane2005,Shevtsov2012} for graphene, a staggered intrinsic SOC does not preserve the gap in the presence of an external magnetic field. The bulk band gap mainly forms between nonzero Landau levels from the $K$ and $K^{\prime}$ valley, which can be shifted by the magnetic field. On the other hand, in the Kane-Mele model, a bulk band gap forms between the zero Landau levels, which is stable under an external magnetic field. In the systems with valley-Zeeman spin-orbit coupling, the bulk band gap closes and reopens at a crossover value of the external magnetic field equal to \cite{Frank2020}
\begin{align}
    B_{c}=
    \frac{
    \left[\lambda_I^A-\lambda_I^B-2\Delta\right]
    \left[\left(\Delta+\lambda_I^A\right)\left(\Delta-\lambda_I^B\right)+4\lambda_R^2\right]
    }
    {
    2e\hbar v_F^2\left[2\Delta+\lambda_I^A-\lambda_I^B\right]
    }.
\end{align}
For the given parameters set from Ref.~\cite{Gmitra2015,Frank2018} of our Hamiltonian\eqref{eq:tb-ham}, the crossover value of the magnetic field will be equal to 1.942~mT for the graphene/WSe$_2$ heterostructure. This crossover point distinguishes two regimes, below the crossover field, a bulk band gap formed by the different nonzero Landau levels, while above the crossover field bounded by the first Landau levels.

\subsection{Zigzag ribbon results}

\begin{figure}
\begin{center}
\includegraphics[width=.5\textwidth]{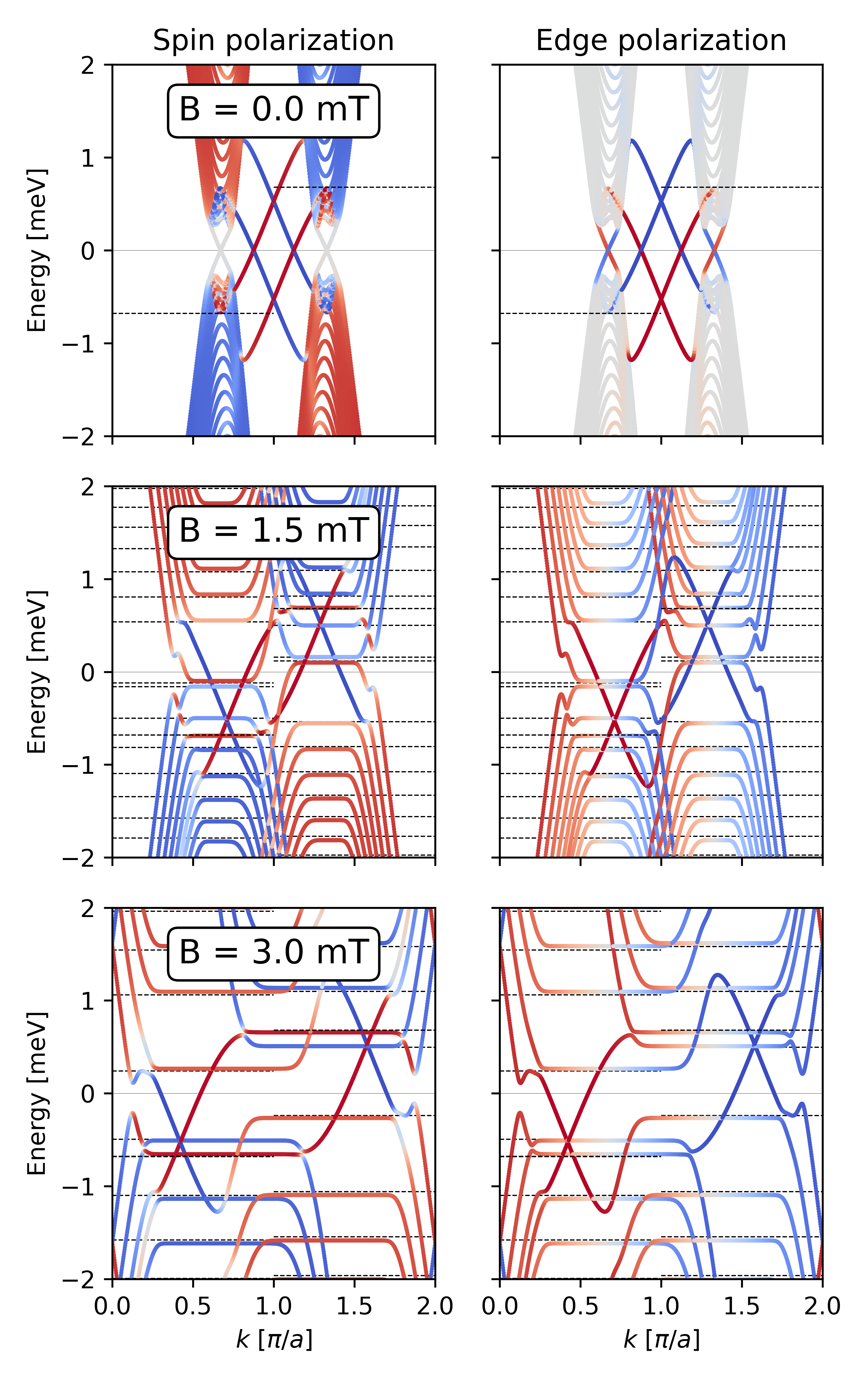}
\end{center}
\label{fig:band}
\caption{Calculated band structures of a proximitized graphene zigzag ribbon in the presence of a perpendicular magnetic field B=0~mT (a), B=1.5~mT, and B=3~mT. The color code denotes spin polarization in the left column and edge polarization in the right column. Thin dashed lines indicate bulk Landau levels.}
\end{figure}

The band structure of a zigzag graphene ribbon is calculated for different values of the external perpendicular magnetic field. We consider zigzag ribbons
with the width of 4.1~$\mu$m. With the scaling factor $r=400$, the simulation size of the ribbon is reduced to  72 carbon-carbon bonds. The results for the zigzag ribbon band structure for magnetic fields of 0~mT, 1.5~mT, and 3.0~mT are shown in Fig.~\ref{fig:band}. At zero magnetic field, we can distinguish two types of edge states: spin unpolarized intravalley states and pseudohelical intervalley states with strong spin polarization~\cite{Frank2018, Gmitra2016}. The intravalley edge states are more delocalized towards the bulk than intervalley states, due to their spectral proximity to the bulk states. 

To compare with the bulk~\cite{Frank2020}, we also plot bulk Landau levels at $K$ and $K^{\prime}$. For small magnetic fields (1.5~mT), the Landau levels start to form and bands at the $K$ and $K^{\prime}$ valleys lose their dispersion. In general, the Landau levels are rather bulk-like, indicated by the absence of edge polarization. They coincide with the analytic prediction\cite{Frank2020}. 

At the borders of the bulk continuum, edge states form, which are responsible for the quantum Hall effect in graphene. With the increase of the external magnetic field, intervalley edge states shift in $k$-space: states with the same spin polarization shift in the same direction, while states with the opposite spin polarization shift in the opposite direction. Generally, intervalley edge states are not much affected by the perpendicular magnetic field. However, the intravalley states get strongly modified,  merging with the bulk bands; this happens already at magnetic fields less than 1.5~mT. This is similar to what is predicted for 2D topological insulators in a magnetic field~\cite{Bottcher2019}. The bulk gap is openning at the crossover magnetic field $B_c \approx  1.942$~mT. As 
a consequence, the intravalley edge states disappear, leaving behind the lone pairs of pseudohelical edge states, ona pair at each zigzag edge. 

\subsection{Flake results}
To analyze the nature of the edge states, we have calculated low energy states of differently sized proximitized graphene flakes. At zero magnetic field the pseudohelical states which form at the zigzag edges reflect back to the intravalley states, which generates standing waves spread out through the flake. In nanoribbons, with sizes less than a micron, the intravalley states are gapped out and the pseudohelical states tunnel through the armchair edges, allowing them to fully propagate through the flake edges. 

\begin{figure}
\begin{center}
\includegraphics[width=.5\textwidth]{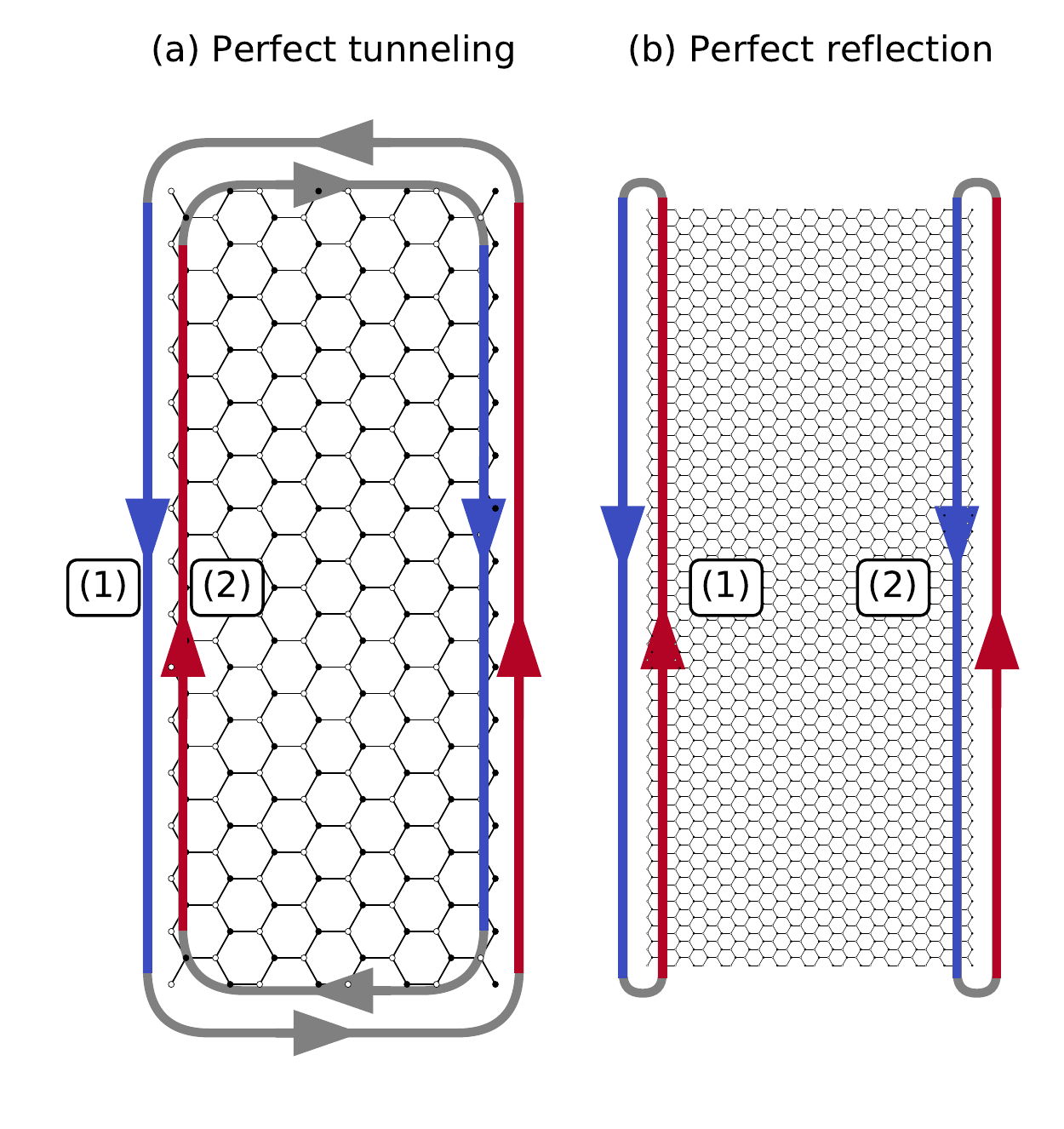}
\end{center}
\caption{(a) Schematic representation of pseudohelical states in a proximitized graphene flake. The states form at zigzag edges, propagate towards the armchair edges, along which they tunnel and undergo spin flip due to Rashba coupling. The colors represent the spin which is perpendicular to the sheet. At zero magnetic field the tunneling along the armchair edges is perfect. (b) In large flakes, if $B > B_c$, the counterpropagating pseudohelical edge states are perfectly reflected at the armchair edges. The pair then forms a standing, nonpropagating wave, which carries net spin current, but no charge current. The resulting state is spin unpolarized. See also Fig.~\ref{fig:schema}(c).}
\label{fig:states}
\end{figure}

As we showed above, for $B > B_c$ the intravalley states in wide ribbons are gapped out as well. Can then the pseudohelical states follow the same scenario as in nanoribbons? The answer is no. While at zero magnetic field the pseudohelical states cannot reflect back at the armchair edges due to time-reversal symmetry (the pseudohelical pair is formed by time-reversal partners), in the presence of a magnetic field such a reflection is possible. In fact, as we demonstrate below, this reflection is perfect, leading to the localization of the pseudohelical pair at zigzag edges and formation of standing waves that carry pure spin current. 
The qualitative difference between propagating pseudohelical edge states and pure spin current non-propagating states is depicted in Fig.\ref{fig:states}.

While the formation of the pseudohelical states can be explained by perfect tunnelling of the intervalley states across the armchair edge, the formation of the pure spin current states corresponds to perfect reflection off the armchair edge. Such a difference in the mechanisms of the formation of states is possible due to the exponential decay of the probability of tunnelling between zigzag edges as the distance between them increases. It is not possible to continuously go from one limit to another due to the periodic subband gap opening and closing~\cite{Frank2018}. However, we can clearly distinguish pseudohelical states for a narrow graphene flake and pure spin current states for a wide graphene flake, because these states represent two different physical limits. 
We estimated the distance between zigzag edges at which perfect tunneling still occurs as the typical spin-flip length of the Rashba spin-orbit interaction
\begin{align}
    l_{R}\approx3a\frac{t}{\lambda_R}.
\end{align}
For our parameters (graphene/WSe$_2$) $l_{R}$ is about 3.4~$\mu$m, which is less than the flake width we use in this work. 

Pseudohelical states and pure spin current states are formed by combinations of intervalley states. They both consist of states (1) and (2) at the left edge and of (3) and (4) at the right edge, see  Fig.\ref{fig:states}(c). Both types of states, pseudohelical states and pure spin current states, are formed using the same edge states and the only difference is the mechanism of their combination.

We have specifically investigated pure spin current (i.e., spin current without any charge current) states for a flake whose size is equal to 16.3584~$\mu$m $\times$ 4.0896~$\mu$m, taking into account the scaling factor $r=400$, which corresponds to an auxiliary 288 $\times$ 72 carbon-carbon bond sized flake. The width of this flake is greater than the Rashba spin-flip length, so the perfect reflection mechanism is dominant in this case. We trace the evolution of the highest occupied state with incremental steps in the external perpendicular magnetic field. 

\begin{figure}
\begin{center}
\includegraphics[width=.5\textwidth]{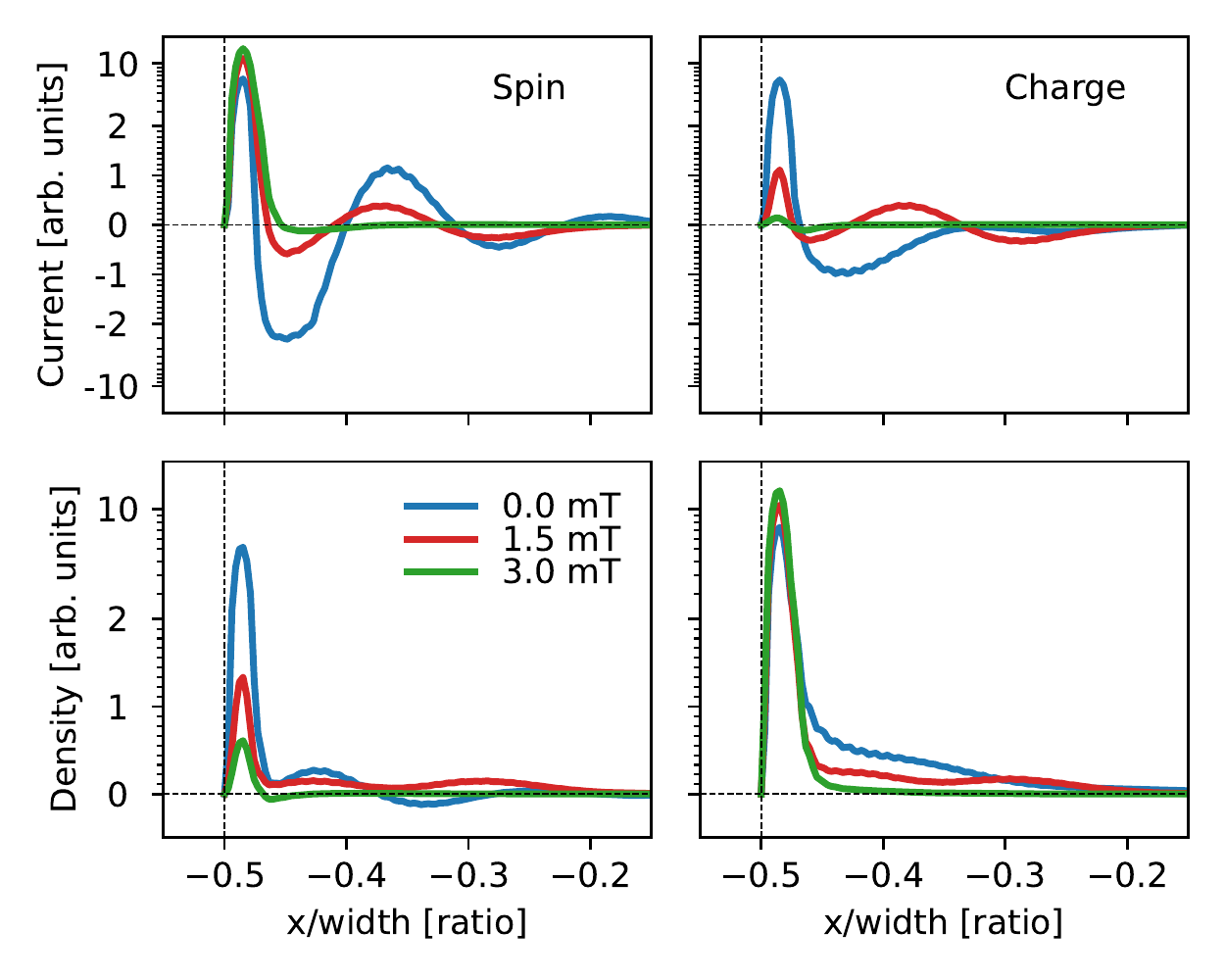}
\end{center}
\caption{Calculated spin (a) and charge (b) currents of the highest occupied flake states (see text for flake parameters) through the cut in the middle of the flake and in the presence of a
perpendicular magnetic field. Spin (c) and charge (d) densities of the highest occupied state at the cut in the middle of the flake and in the presence of a magnetic field.}
\label{fig:cut}
\end{figure}

The spin and charge currents and densities through a cut in the middle of the flake, orthogonal to the armchair direction are shown in Fig.\ref{fig:cut}. The data indicates that with the increase of the magnetic field, the charge current and spin densities disappear and complete edge localization of the state occurs. Also, the spatial spin current oscillations vanish and the highest occupied states become pure spin current states with negligible spin density. A similar situation occurs with the lowest unoccupied state. 

\begin{figure}
\begin{center}
\includegraphics[width=.5\textwidth]{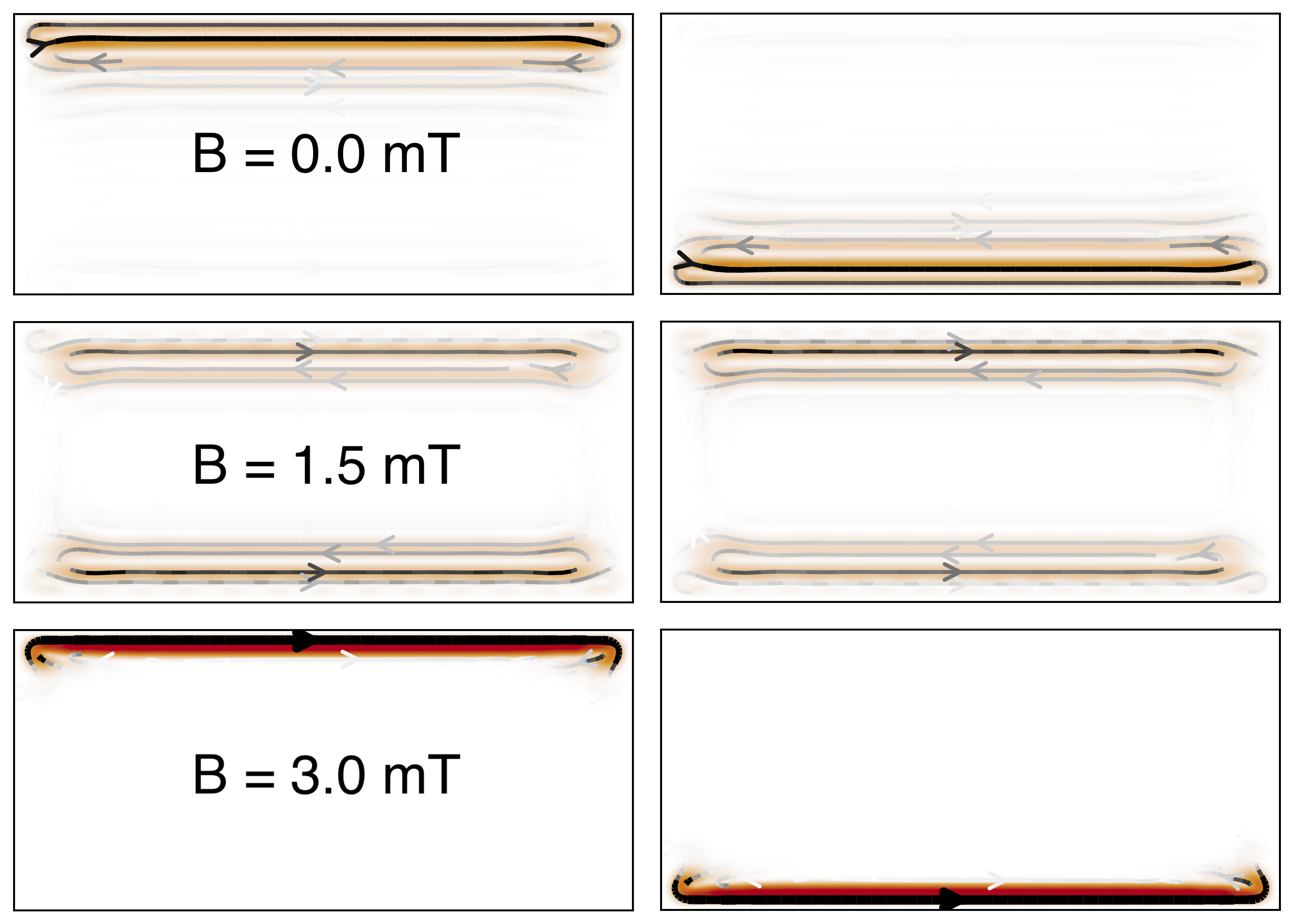}
\end{center}
\caption{Evolution of the spin current of the highest occupied flake state in the left column and of the lowest unoccupied flake state in the right column.}
\label{fig:spin}
\end{figure}

\begin{figure}
\begin{center}
\includegraphics[width=.5\textwidth]{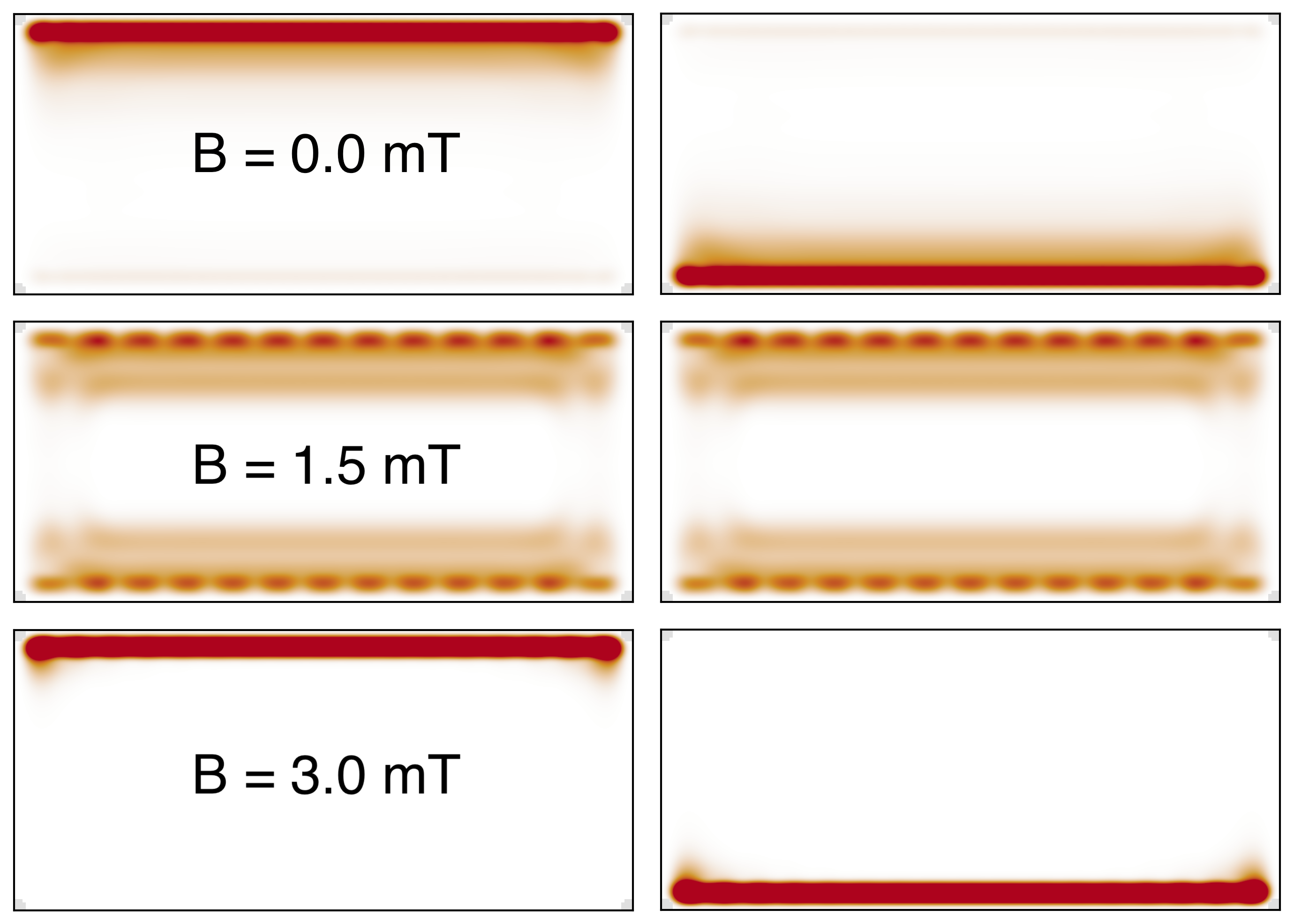}
\end{center}
\caption{Evolution of the charge density of the highest occupied flake state in the left column and of the lowest unoccupied flake state in the right column.}
\label{fig:charge}
\end{figure}

The evolution of the spin currents of the highest occupied and the lowest unoccupied states are shown in Fig.\ref{fig:spin}. Both states exhibit similar behavior of their spin currents, located on different zigzag edges of the flake. Pure spin currents arise in the presence of an external magnetic above the crossover value $B_c$. The evolution of the charge density is similar to the previous case, depicted in Fig.~\ref{fig:charge}. With an external magnetic field below the crossover value, the states closest to the Fermi level are a mixture of strongly localized intervalley states and weakly localized intravalley states; they form a standing wave at the zigzag boundary, as tunneling through the armchair edge is forbidden (this can explicitly be seen in gapped ribbon armchair band structures~\cite{Frank2018}).


However, with a magnetic field above the crossover value $B_c$, we find states with pronounced pure spin current, localized strongly at a single edge. From the zigzag ribbon band-structure analysis, we can conclude that the pure spin current states correspond to the combination of the intervalley edge states on the same edge, which is stable under the application of a magnetic field. Similar to the pseudohelical states, pure spin current states are stable against scattering on defects at the zigzag edge of the graphene flake, see Appendix A.

\section{Conclusion}
We investigated the behavior of the electronic structure of proximitized graphene ribbons and flakes of microscopic sized in the presence of a perpendicular magnetic field. We have chosen 
graphene/WSe$_2$ heterostructure for the numerical parameters, as in such structures pseudohelical states were predicted to exist. We found that under the influence of a magnetic field, the pseudohelical (intervalley) edge states are preserved, while the intravalley edge states disappear at the crossover value of the magnetic field $B_c \approx 1.9~mT$. This value corresponds to the closing/reopening of the gap between the bulk Landau levels. 

We also studied a finite flake of micron sizes, 
finding that instead of perfect tunneling of the pseudohelical states at the armchair edges at zero magnetic field, for $B > B_c$ the states perfectly reflect to their counterpropagating partners and form non-propagating, spin unpolarized pure spin current states which are stable under scattering off zigzag edge defects. 

Pure spin current states should be observable in wide flakes of at least a few microns (for graphene/TMDC heterostructures) already at rather weak (a few millitesla) fields. Lone pseudohelical pairs at a given edge should be observable in wide ribbons for $B > B_c$ of in nanosized flakes where perfect tunneling through armchair edges is allowed and intravalley states are gapped even at zero magnetic field.

\begin{acknowledgments}
We would like to thank D. Kochan for useful discussions. This work was supported by the DFG SFB Grant No. 1277 (A09 and B07), DFG SPP 2244 (Project-ID 443416183), the International Doctorate Program Topological Insulators of the Elite Network of Bavaria. The authors gratefully acknowledge the Gauss Centre for Supercomputing e.V. (www.gauss-centre.eu) for providing computing time on the GCS Supercomputer SuperMUC at Leibniz Supercomputing Centre (LRZ, www.lrz.de).
\end{acknowledgments}

\appendix

\begin{figure}
\begin{center}
\includegraphics[width=.5\textwidth]{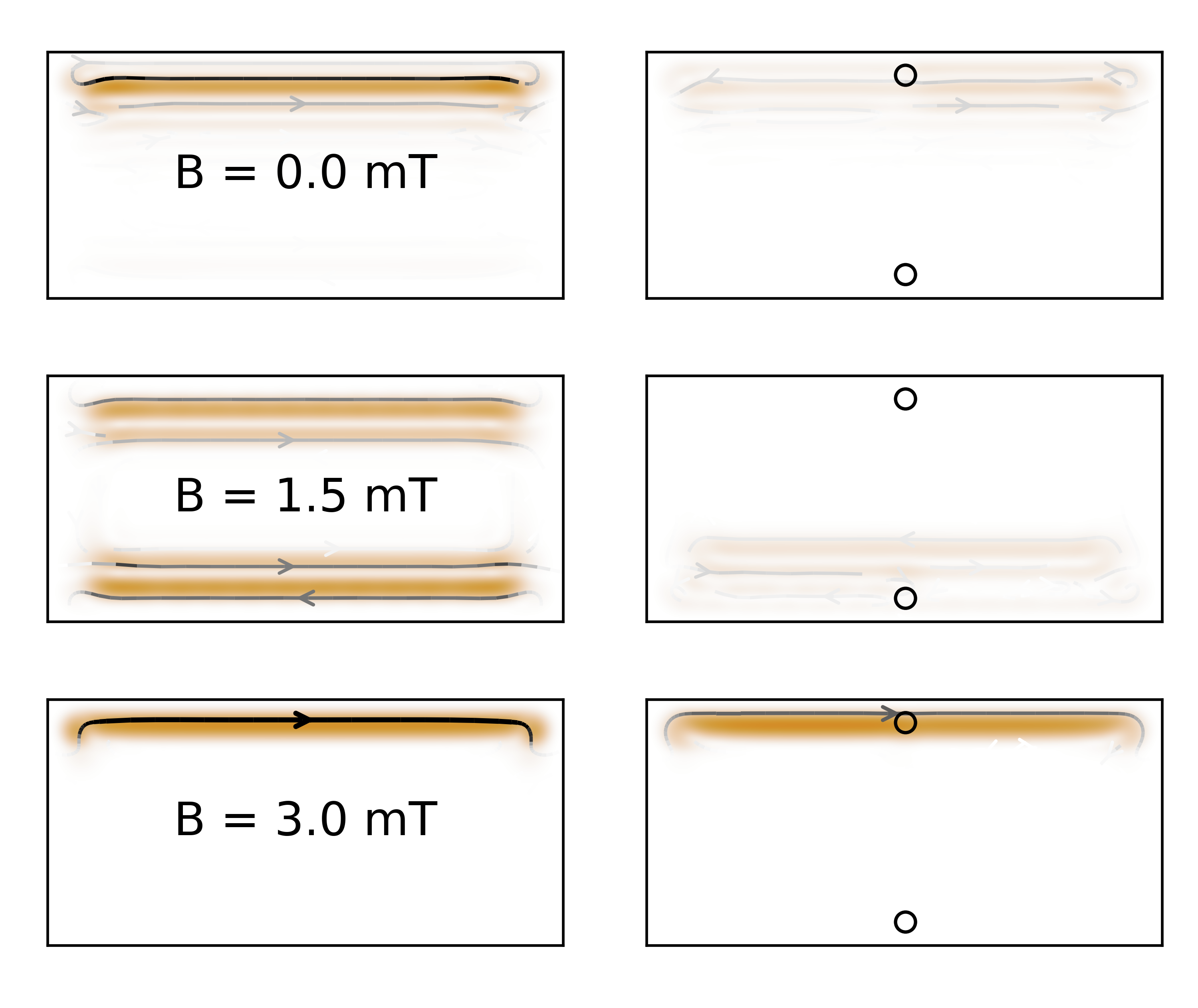}
\end{center}
\caption{Evolution of the spin current of the highest occupied flake state without defects in the left column and with two defects on zigzag edges in the right column. Edge defects are represented as circles. Spin current in the pure spin current regime is stable under the influence of the internal scatters.}
\label{fig:spin-defect}
\end{figure}

\begin{figure}
\begin{center}
\includegraphics[width=.5\textwidth]{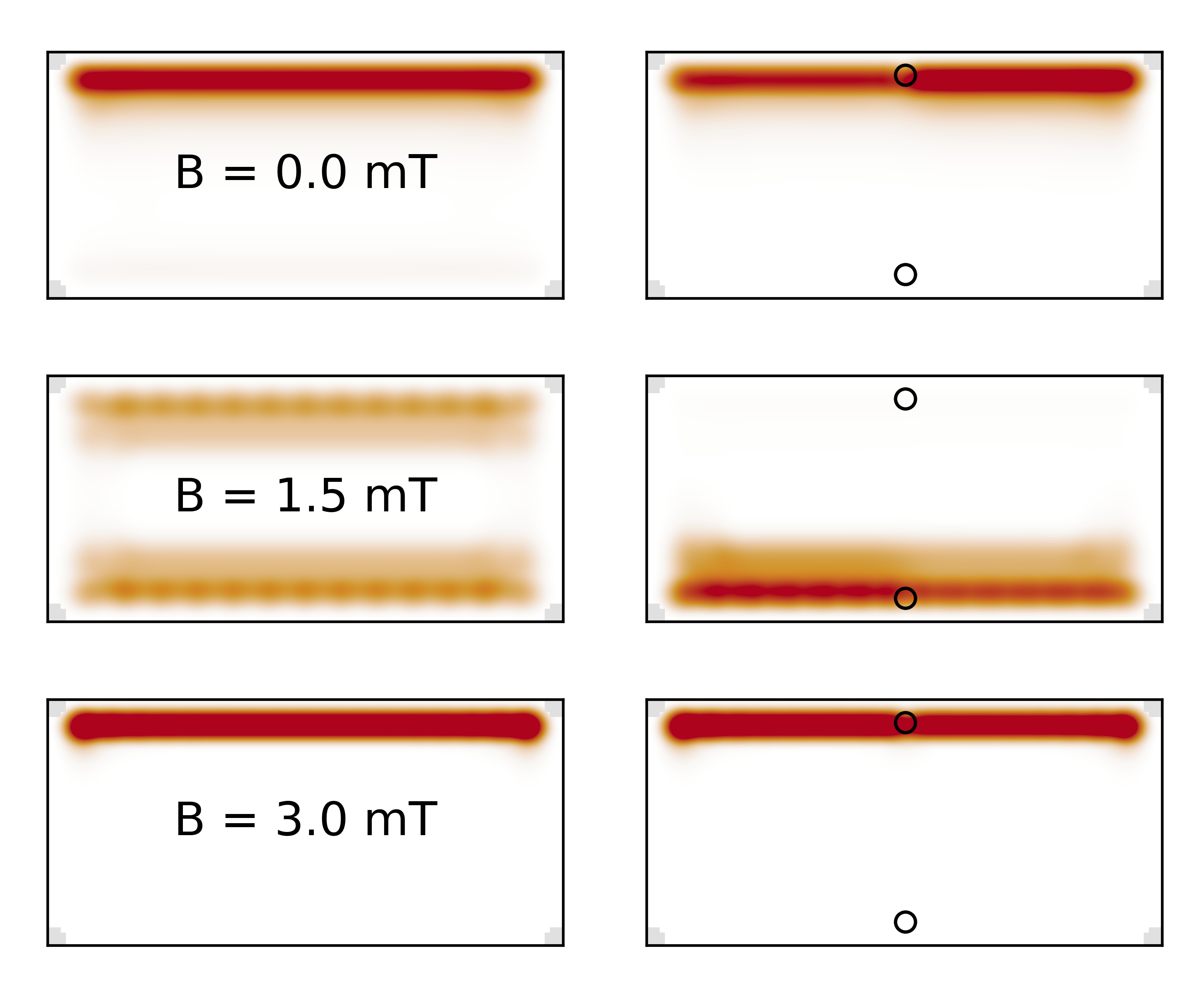}
\end{center}
\caption{Evolution of the charge density of the highest occupied flake state without defects on the left column and with two defects on zigzag edges on the right column. Internal defects are represented as circles. Charge density in the pure spin current regime is stable under the influence of the internal scatters.}
\label{fig:charge-defect}
\end{figure}

\section{Scattering on defects}
To demonstrate the stability of pure spin current states under the action of internal scatterers, we calculate the spin current and the charge density in the presence of two defects on the zigzag edges of a proximitized graphene flake. The calculation results are presented in Figs. \ref{fig:spin-defect} and \ref{fig:charge-defect}. At magnetic field values above the crossover $B_c$, the spin current and charge density remain unchanged after adding internal defects to the zigzag edges by removal of a lattice site.

\end{document}